\documentclass[%
 aip,reprint, jcp, floatfix
 reprint,%
]{revtex4-1}

\usepackage{graphicx}
\usepackage{dcolumn}
\usepackage{bm}

\usepackage{gensymb} 
\usepackage{float}

\usepackage[utf8]{inputenc}
\usepackage[T1]{fontenc}
\usepackage{mathptmx}
\usepackage{etoolbox}

\usepackage[version=3]{mhchem} 
\usepackage{soul}
\usepackage{subcaption}
\usepackage{xcolor}
\usepackage{multirow}
\usepackage{hyperref}
\hypersetup{
    colorlinks=true,
    linkcolor=blue,
    urlcolor=cyan
}

\makeatletter
\def\@email#1#2{%
 \endgroup
 \patchcmd{\titleblock@produce}
  {\frontmatter@RRAPformat}
  {\frontmatter@RRAPformat{\produce@RRAP{*#1\href{mailto:#2}{#2}}}\frontmatter@RRAPformat}
  {}{}
}%
\makeatother
\begin{document}

\preprint{JCP}

\title[Microsecond-scale sucrose conformational dynamics in aqueous solution via molecular dynamics methods]{Microsecond-scale sucrose conformational dynamics in aqueous solution via molecular dynamics methods}
\author{Vladimir I. Deshchenya}
\email{deshchenia.vi@phystech.edu}
\affiliation{Center for Computational Physics, Moscow Institute of Physics and Technology (National Research University), Institutskiy Pereulok 9, Dolgoprudny, 141701, Moscow oblast, Russia}%
\affiliation{Joint Institute for High Temperatures RAS, Izhorskaya st. 13 Bd. 2, Moscow, 125412, Russia}
\author{Kirill M. Gerke}
\affiliation{Center for Computational Physics, Moscow Institute of Physics and Technology (National Research University), Institutskiy Pereulok 9, Dolgoprudny, 141701, Moscow oblast, Russia}%
\affiliation{Schmidt Institute of Physics of the Earth RAS, Bolshaya Gruzinskaya st. 10-1, Moscow, 107031, Russia}
\author{Nikolay D. Kondratyuk}
\affiliation{Center for Computational Physics, Moscow Institute of Physics and Technology (National Research University), Institutskiy Pereulok 9, Dolgoprudny, 141701, Moscow oblast, Russia}%
\affiliation{Joint Institute for High Temperatures RAS, Izhorskaya st. 13 Bd. 2, Moscow, 125412, Russia}
\affiliation{HSE University, Myasnitskaya st. 20, Moscow, 101000, Russia}

\date{\today}

\begin{abstract}
Molecular dynamics methods have proven their applicability for the reproduction and prediction of molecular conformations during the last decades. However, most of works considered dilute solutions and relatively short trajectories that limit insights into conformational dynamics. In this study, we investigate the conformational dynamics of sucrose in aqueous solution using microsecond-scale molecular dynamics simulations. For the most of the calculations we use the OPLS-AA/1.14*CM1A-LBCC force field, but we also utilize OPLS-AA/1.14*CM1A and GLYCAM06 for the comparison. We focused on the glycosidic linkage conformers and their lifetimes, glucopyranose and fructofuranose ring puckering. Our findings indicate that the $^1\mathrm{C}_4$ glucopyranose ring conformation can stabilize the sucrose conformer, appeared only in the GLYCAM06 and OPLS-AA/1.14*CM1A force fields. All the results are strengthened by comparison with the available experimental data on NMR J-coupling constants and ultrasonic spectra.

\textit{This article may be downloaded for personal use only. Any other use requires prior permission of the author and AIP Publishing. This article appeared in J. Chem. Phys. 163, 044502 (2025) and may be found at https://doi.org/10.1063/5.0266322. © 2025 Author(s). This article is distributed under a Creative Commons Attribution-NonCommercial-NoDerivs 4.0 International (CC BY-NC-ND) License.}
\end{abstract}

\maketitle

\section{Introduction}
Carbohydrates are fundamental to numerous biological processes, and their functions are critically dependent on their conformational behavior. A detailed understanding of their spatial arrangements is key to elucidating their stereochemistry, biological roles~\cite{takeda2024viscosity}, and interactions with proteins and other biomolecules~\cite{Xiuming2015, Calabrese2019, Yang2020, Toukach1, Toukach2}.
Such knowledge enables the design of drugs~\cite{khrabrov2022nabladft,khrabrov2024nabla} that precisely target carbohydrate-binding sites, and optimize the performance of carbohydrate-based biomaterials.
In most natural glycans, which are predominantly composed of pyranose moieties, conformational behavior involves relatively stable monosaccharides occupying one or more predominant conformations and flexible bonds between them~\cite{hricovni2004structural}. As the simplest representatives of this class, disaccharides embody the full range of rotational degrees of freedom found in oligo- and polysaccharides~\cite{stroylov2020comparison}. 

The conformational dynamics of sucrose, one of the most abundant and biologically significant disaccharides, in water have been the focus of several studies~\cite{Imberti2019, Olsson2020, Silva2022}. Nuclear magnetic resonance (NMR) spectroscopy has been widely employed to demonstrate that sucrose molecule has multiple conformers in aqueous solution~\cite{duker1993, freedberg2002, Venable2005}. Moreover, Kaatze and colleagues analyzed the ultrasonic spectra of mono- and disaccharide solutions, identifying a disaccharide-specific relaxation process related to ring rotations (i.e., changes in glycosidic linkage conformation), its characteristic timescale, and its dependence on concentration~\cite{kaatze1, kaatze2}. Despite these advances, the conformational dynamics of sucrose in water remain incompletely understood. Discrepancies persist regarding the number of conformers present, and critical properties such as their lifetimes have not been thoroughly characterized~\cite{bock1982, immel1995molecular, xia2011sucrose1}.

Molecular dynamics (MD) simulations have emerged as a powerful tool for investigating the conformational dynamics of molecules in solvents~\cite{chekmarev2004long,kwac2008classical,husic2018markov,gurina2024self, gaur2022conformer, Medvedev2025, Xu2025, Koneru2019, Kolarikova, Lutsyk2024, Dorst2024}. They provide detailed insights into the trajectories of atoms and molecules over time, allowing to analyze transitions between conformers and stability in a solvent environment, combined with the statistical models~\cite{fedotova2020hydration,chuev2022molecular,kruchinin2023silico}. However, the system’s evolution is described by the interatomic potential, and the reliability of the results depends on the suitability of the force field~\cite{vitalini2015dynamic,sweere2017accuracy,bakulin_properties_2021,kadaoluwa2021systematic, volkov2021molecular}. 

Finding a suitable force field for carbohydrates in aqueous solutions has been a long-standing challenge~\cite{batista2015evaluation, lay2016optimizing, jamali2018optimizing, deshchenya2023history}. A comprehensive review on the applicability of various force fields in terms of hydration properties, conformational analysis and diffusion is published very recently~\cite{Deegbey}. Early generic and carbohydrate-specific force fields, such as OPLS-AA, GLYCAM06~\cite{glycam06}, CHARMM36, and GROMOS 56A$_\textrm{carbo}$ lead to overaggregation of carbohydrate molecules in water and incorrect dynamic properties of solution. However, reparameterization of intramolecular nonbonded interactions for specific carbohydrates has been found to be a solution to these problems~\cite{batista2015evaluation, lay2016optimizing, jamali2018optimizing, Prado2023, Pluharova}.
Also, it has been demonstrated that the OPLS-AA force field with partial charge correction 1.14*CM1A~\cite{dodda20171, dodda2017ligpargen} accurately reproduces both the dynamical properties and solution behavior~\cite{deshchenya2022molecular, deshchenya2023history}.

To date, the most comprehensive MD investigation of sucrose was conducted by Xia and Case~\cite{xia2011sucrose1, xia2011sucrose2}, who employed both classical and accelerated MD simulations to explore its conformation dynamics in aqueous and dimethyl sulfoxide solutions. Their first work examined a dilute solution containing a single sucrose molecule, while a subsequent work addressed finite concentrations and concluded that the sucrose conformer distribution is independent of concentration. However, these investigations utilized the GLYCAM06 force field, which has been shown to lead to overaggregation of sucrose molecules, and they did not consider this issue.

In our previous study validating the OPLS-AA/1.14*CM1A~\cite{dodda20171, dodda2017ligpargen} force field for sucrose in aqueous solution~\cite{deshchenya2022molecular}, we also examined the conformational preferences of sucrose molecule. Consistent with the findings of Xia and Case, we identified four distinct conformers. However, the relatively short trajectory lengths of 10~ns did not allow us to investigate the lifetimes of these conformations. Furthermore, the present study revealed that the OPLS-AA/1.14*CM1A force field results in an erroneous distribution of glucopyranose ring conformations at longer timescales. {\color{black}Here, we show that the it leads to the appearance of the fourth conformer previously obtained for OPLS force fields.}

{\color{black}In this study, we utilized microsecond-scale molecular dynamics simulations to investigate the conformational behavior of sucrose in aqueous solution across three concentrations (20\%, 30\%, and 50\% by mass). The trajectories are longer by two orders of magnitude compared to our previous work~\cite{deshchenya2022molecular}. Analyzing this long dynamics, we focused on the J-coupling constants, glycosidic linkage conformers, their lifetimes, glucopyranose and fructofuranose rings puckering, comparing the OPLS-AA/1.14*CM1A, OPLS-AA/1.14*CM1A-LBCC, and GLYCAM06 force fields. We were able to compare J-coupling constants and conformers lifetimes with the experimental data. We addressed the overaggregation tendencies observed with GLYCAM06 and examined discrepancies previously noted in the literature.}

\section{Modeling and simulation techniques}
\subsection{Interatomic Potential} 

The idea behind the molecular dynamics method is to numerically integrate Newton's equation of motion for each of the particles of the system and then analyze the resulting trajectory using statistical physics methods. 

The interatomic interaction potential $U(\boldsymbol{r}_1, ..., \boldsymbol{r}_N)$ determines the accuracy of the property predictions. In this paper, the modern version of Optimized Potentials for Liquid Simulations All-Atom (OPLS--AA) force field with partial charge correction 1.14*CM1A {\color{black}and localized bond-charge correction(LBCC)~\cite{dodda20171, dodda2017ligpargen} was used. This correction (redistribution of charge) was done by Dodda et al. systematically for several functional groups. The procedure was then validated by the reproduction of hydration energies and pure liquids densities.} For water we used the four-point model TIP4P/2005~\cite{lfabascal2005general}.

Previously we extensively validated the combination of OPLS-AA/1.14*CM1A \textit{without} LBCC force field and TIP4P/2005 water. This combination eliminates the problem of overaggregation of sugar molecules in aqueous solution and yields good agreement between calculated and experimental densities and transport coefficients of sucrose aqueous solution over a wide range of sugar concentrations and temperatures~\cite{deshchenya2022molecular}. Here, we checked that OPLS-AA/1.14*CM1A-LBCC in combination with TIP4P/2005 reproduces similar density and diffusion coefficients as OPLS-AA/1.14*CM1A and also does not lead to overaggregation.

Also, calculations with the GLYCAM06 force field~\cite{glycam06} were performed for one solution, with 50\% mass fraction of sucrose, for comparison, as well as OPLS-AA/1.14*CM1A \textit{without} LBCC.

\subsection{Molecular Dynamics details}
Studying conformational transitions requires computationally intensive simulations on the microsecond-scale, where the choice of simulation software and parameters, including system size, is particularly important~\cite{kutzner2019more, kutzner2022gromacs, kondratyuk2021gpuaccelerated}. All MD simulations in this work were performed using OpenMM~\cite{eastman2023openmm}, an open-source Python library for biomolecular simulations. It is well-optimized for the GPU-based calculations, with the central processing unit used solely for data exchange.

This study examined sucrose solutions with mass fractions ($w_s$) of 20\%, 30\%, and 50\% to investigate the concentration dependence. All the simulation boxes contained 200 sucrose molecules and 15200 ($w_s$ = 20\%), 8867 ($w_s$ = 30\%), and 3800 ($w_s$ = 50\%) water molecules. These system sizes were chosen to ensure optimal computational performance~\cite{deshchenya2023history}. All simulations were conducted at 293~K and 1~bar.

\begin{figure}
\centering
\includegraphics[width=1\linewidth]{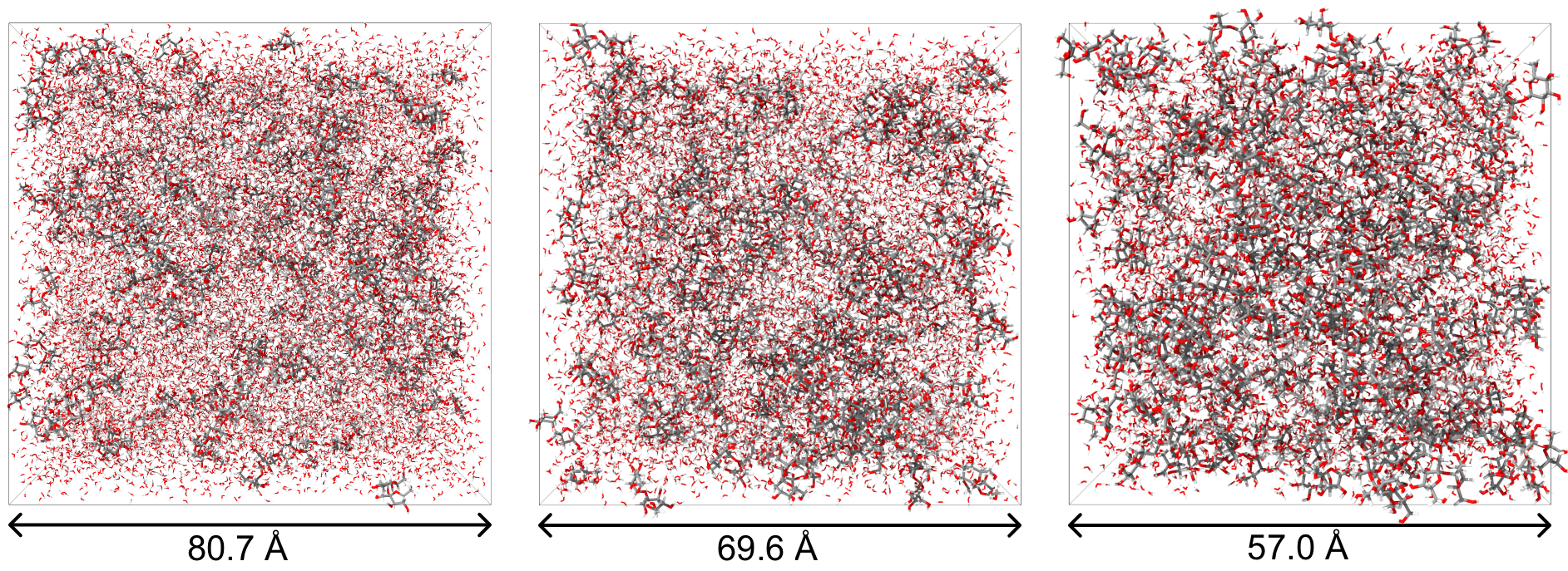}
\caption{Visualizations of simulation boxes of aqueous sucrose solutions for three concentrations: 20\% (left), 30\% (center), and 50\% (right). Visualization was done using the ChimeraX software~\cite{pettersen2020ucsf}.}
\label{pic:simboxes}
\end{figure}

The initial molecular arrangements were generated using the Packmol package~\cite{martinez2009packmol}, with sucrose in its crystal conformation~\cite{hanson1973sucrose}. Figure~\ref{pic:simboxes} shows the simulation boxes visualizations for three concentrations (20\%, 30\%, and 50\%).

In all simulations, a consistent equilibration protocol was implemented. The first step involved energy minimization of the system using the L-BFGS algorithm~\cite{liu1989limited}, after which atomic velocities were assigned according to a Maxwell distribution for the desired temperature. Next, molecular dynamics simulations were initiated in the canonical~(NVT) ensemble to stabilize the temperature, followed by equilibration in the isothermal-isobaric~(NPT) ensemble to calculate the equilibrium density. For the OPLS-AA/1.14*CM1A-LBCC force field, the equilibrium densities were 1.079~g/cm$^3$~(for 20\% solution), 1.124~g/cm$^3$~(for 30\%), and 1.225~g/cm$^3$~(for 50\%), while for the GLYCAM06 force field, the equilibrium density was 1.226~g/cm$^3$ for the 50\% solution. Further simulations and calculations of target properties were done in the canonical~(NVT) ensemble at this equilibrium density. The Nosé-Hoover thermostat~\cite{nose1984, hoover1985canonical} maintained the temperature, and the Monte Carlo barostat~\cite{chow1995, johan2004molecular} controlled atmospheric pressure during equilibrium NPT runs.

A timestep of 2~fs was used in all simulations. All bonds involving hydrogen atoms were constrained using the SHAKE algorithm. Periodic boundary conditions were applied to avoid surface effects. The Lennard-Jones and electrostatic interaction potentials were truncated at a distance of 12~\AA. Additional corrections to pressure and energy calculations were applied to account for the truncation of potentials~\cite{sun1998compass}. The long-range part of the Coulomb potential was calculated using the Particle Mesh Ewald~(PME) algorithm~\cite{essmann1995smooth}.

\subsection{NMR J-coupling constants calculations}
J-coupling constants provide a reliable and widely used metric for comparing MD simulation results with experimental data, as they are measurable by NMR spectroscopy. They are influenced by molecular geometry and can also be derived from MD trajectories. Vicinal NMR coupling constants are obtained from dihedral angles using the Karplus equation~\cite{Coxon2009}:
\begin{equation}\label{eq:karplus}
J = A \cos^2(\phi) + B \cos(\phi) + C, 
\end{equation}
where $\phi$ is the dihedral angle, and A, B, and C are Karplus parameters, determined by fitting the equation to quantum mechanical calculation or experimental data. These parameters are specific to the chemical group and are valid only for chemically similar environments. {\color{black}To improve accuracy, particularly in substituted systems, several extended Karplus forms include additional harmonics (e.g.,\ $\cos3\phi$, $\sin3\phi$) and electronegativity corrections~\cite{Houseknecht2003,Altona1994}.}

{\color{black}
In this work we extracted 16 dihedral angles from our MD trajectories and computed the corresponding $^3J$ values using the original Karplus equation and its published extensions. Experimental coupling constant data were compiled from Refs.~\cite{duker1993,Perez1991,Venable2005,freedberg2002} and averaged for each dihedral. Agreement between simulation and experiment was quantified by the mean absolute error (MAE):
\begin{align} 
 \text{MAE} = \frac{1}{N} \sum_{i=1}^{N} \left| J_{\mathrm{sim},i} - J^{\mathrm{avg}}_{\mathrm{exp},i} \right|,
\end{align}
where $J_{\mathrm{sim},i}$ is the value computed from MD in this work for the $i$-th dihedral and $J^{\mathrm{avg}}_{\mathrm{exp},i}$ is the average over all experimentally measured values available for that dihedral. The list of considered dihedral angles, their experimental coupling constant values, and the Karplus‐type equations are provided in Table S1 of the supplementary material.
}

\subsection{Ramachandran plot calculations}

\begin{figure}
\centering
\includegraphics[width=0.75\linewidth]{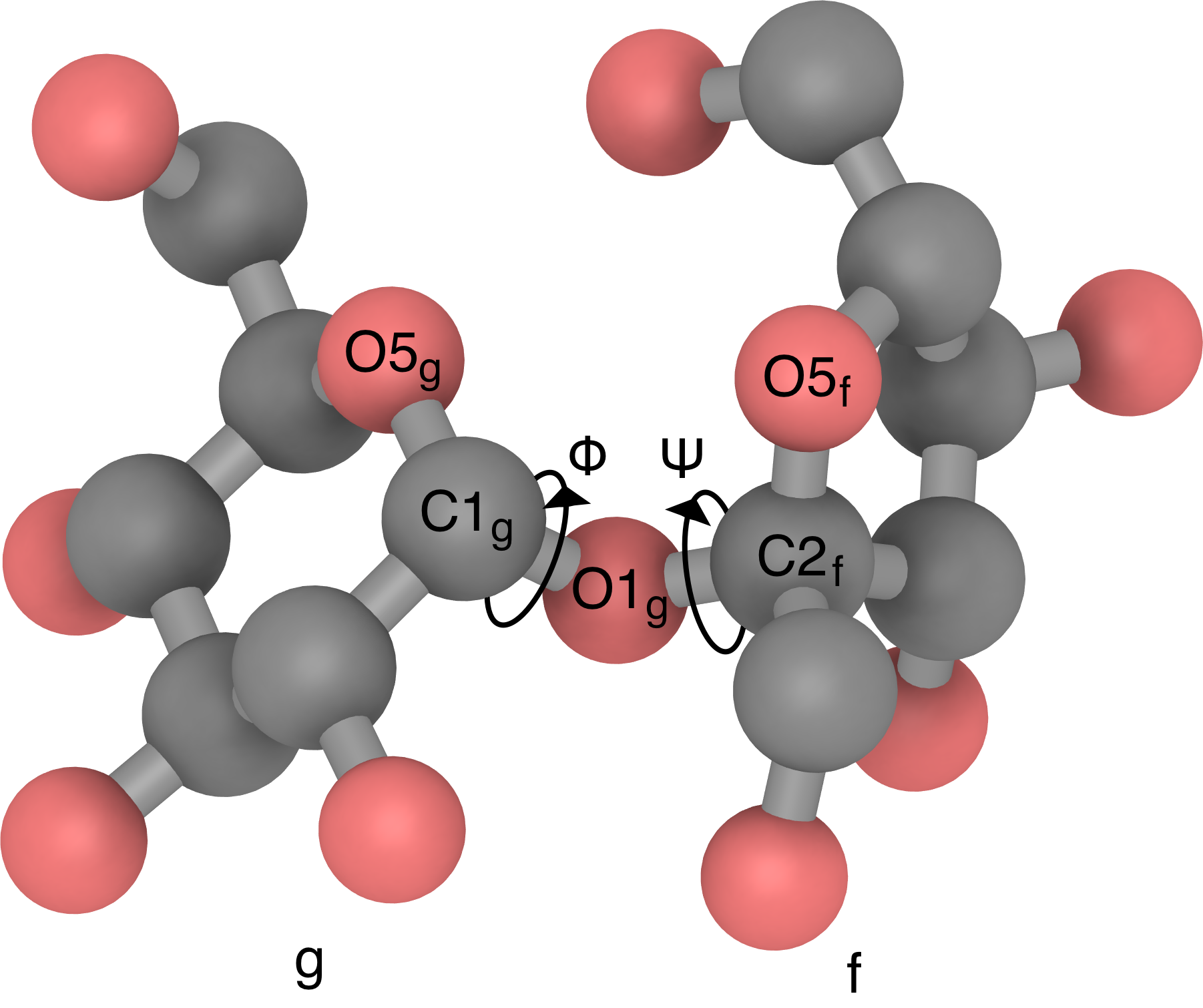}
\caption{Sucrose molecule in its crystalline conformation~\cite{hanson1973sucrose} with dihedral angles $\Phi = \mathrm{O5}_\mathrm{g}$–$\mathrm{C1}_\mathrm{g}$–$\mathrm{O1}_\mathrm{g}$–$\mathrm{C2}_\mathrm{f}$ and $\Psi = \mathrm{C1}_\mathrm{g}$–$\mathrm{O1}_\mathrm{g}$–$\mathrm{C2}_\mathrm{f}$–$\mathrm{O5}_\mathrm{f}$ marked around the glycosidic bond. The letters \textit{g} and \textit{f} denote the glucopyranose and fructofuranose rings of the molecule, respectively. Visualization was done using the iRASPA software~\cite{dubbeldam2018iraspa}.}
\label{pic:sucrosemol}
\end{figure}

The sucrose molecule consists of two rings -- glucopyranose and fructofuranose -- connected by a glycosidic linkage. The molecule conformation is primarily determined by the conformation of this linkage, which is described by two dihedral angles, $\Phi = \mathrm{O5}_\mathrm{g}$–$\mathrm{C1}_\mathrm{g}$–$\mathrm{O1}_\mathrm{g}$–$\mathrm{C2}_\mathrm{f}$ and $\Psi = \mathrm{C1}_\mathrm{g}$–$\mathrm{O1}_\mathrm{g}$–$\mathrm{C2}_\mathrm{f}$–$\mathrm{O5}_\mathrm{f}$. Figure~\ref{pic:sucrosemol} shows the sucrose molecule in its crystalline conformation~\cite{hanson1973sucrose} with labeled atoms and dihedral angles $\Phi$ and $\Psi$ (hydrogen atoms are not shown for clarity).

The joint angular distribution of the dihedrals angles $\Phi$ and $\Psi$ was analyzed using a Ramachandran plot, a two-dimensional free energy surface projected along these dihedral angles. The free energy surface was calculated as
\begin{equation}
\label{freeenergy}
    F(\Phi, \Psi) = -k_B T \ln{P (\Phi, \Psi)},
\end{equation}
where $k_B$ is the Boltzmann constant, $T$ is the absolute temperature of the system, and $P (\Phi, \Psi)$ is the probability of the sucrose molecule adopting a conformation with specific values of the dihedral angles $\Phi$ and $\Psi$. The probability $P (\Phi, \Psi)$ was calculated as the fraction of trajectory within each 1\degree~$\times$~1\degree\ square on a two-dimensional mesh of $\Phi$ and $\Psi$. The global minimum was set to zero by shifting the free energy surface. 

{\color{black}
\subsection{Hydrogen bonds analysis}\label{h_bonds}

The hydrogen bonds could be defined from the geometric or energetic criteria. The first method~\cite{luzar_structure_1993,de_santis_short_1999,ong_temperature-dependent_2019} employs constraints on the spatial arrangement of donor-hydrogen-acceptor atomic groups. The detailed comparison of various geometric criteria could be found in our previous work~\cite{bakulin2024molecular}. The energetic criterion is based on interaction energy thresholds between molecular pairs~\cite{benjamin_structure_1999,sciortino_hydrogen_1989}. These geometric and energy criteria can be applied independently or in combination (e.g.~\cite{chowdhuri_hydrogen_2002}) to refine hydrogen bond definitions. In this work, we employ the geometric definition, which proves adequate for analyzing static properties.

The most widely adopted geometric criterion for hydrogen bonds follows $\angle \text{OHO}^{*} \geq \alpha$, $r_{\text{OO}^{*}} \leq r_1$.
According to this criterion, a hydrogen bond exists when the donor-acceptor ($\text{OO}^{*}$) distances remain below the positions of the first minimum in their respective RDF $r_1$, while the donor-hydrogen-acceptor angle $\angle \text{OHO}^{*}$ stays above $150^\circ$. The acceptable choice for the $r_1$ distance is around 3.4~\AA, which was proven to hold in sucrose aqueous solutions previously~\cite{lerbret2005homogeneous}. We compute hydrogen bond quantities using the MDAnalysis package~\cite{michaud-agrawal_mdanalysis_2011}.}

{\color{black}
\subsection{Conformers' lifetimes calculations}
To estimate the lifetimes of conformers, we employed an approach analogous to that commonly used for hydrogen bond lifetime analysis~\cite{Gowers2015,liu2017hydrogen}. Specifically, we analyzed the time autocorrelation function of the presence of a conformer $C(t)$ and fitted it with a double-exponential function to capture both short- and long-timescale relaxation processes.

For each conformer, the time autocorrelation function was computed as follows:
\begin{equation} \label{eq:C}
    C(t) = \left<\frac{\sum h_i(t_0) h_i (t_0 + t)}{\sum h_i(t_0)^2}\right>,
\end{equation}
where $h_i(t)$ is an indicator function that equals 1 if the $i$-th molecule resides within the specified conformational basin at time $t$, and 0 otherwise. The summation is performed over all sucrose molecules, and the angular brackets denote averaging over all possible initial time points $t_0$.

The resulting autocorrelation function $C(t)$ was then fitted using a biexponential decay model:
\begin{equation} \label{eq:Cfit}
C(t) = A \exp \left(-\frac{t}{\tau_1}\right) + (1-A) \exp \left(-\frac{t}{\tau_2}\right),
\end{equation}
where A, $\tau_1$, and $\tau_2$ are fitting parameters corresponding to the amplitude and relaxation times of the two processes.

The average conformer lifetime, $\tau$, was subsequently calculated as the weighted sum of the two relaxation times:
\begin{equation} \label{eq:tau}
\tau = A \tau_1 + (1-A) \tau_2.
\end{equation}
}

\section{Results and discussion}

One-microsecond trajectories were obtained for each system: sucrose solutions with sucrose mass fractions of 20\%, 30\%, and 50\% for the OPLS-AA/1.14*CM1A-LBCC force field, and a 50\% solution for the GLYCAM06 and OPLS-AA/1.14*CM1A force fields. Dihedral angles $\Phi$ and $\Psi$ for all sucrose molecules were recorded every 200~fs, while the atomic coordinates were saved every 5~ps.

\begin{figure}[h]
\centering
\includegraphics[width=1\linewidth]{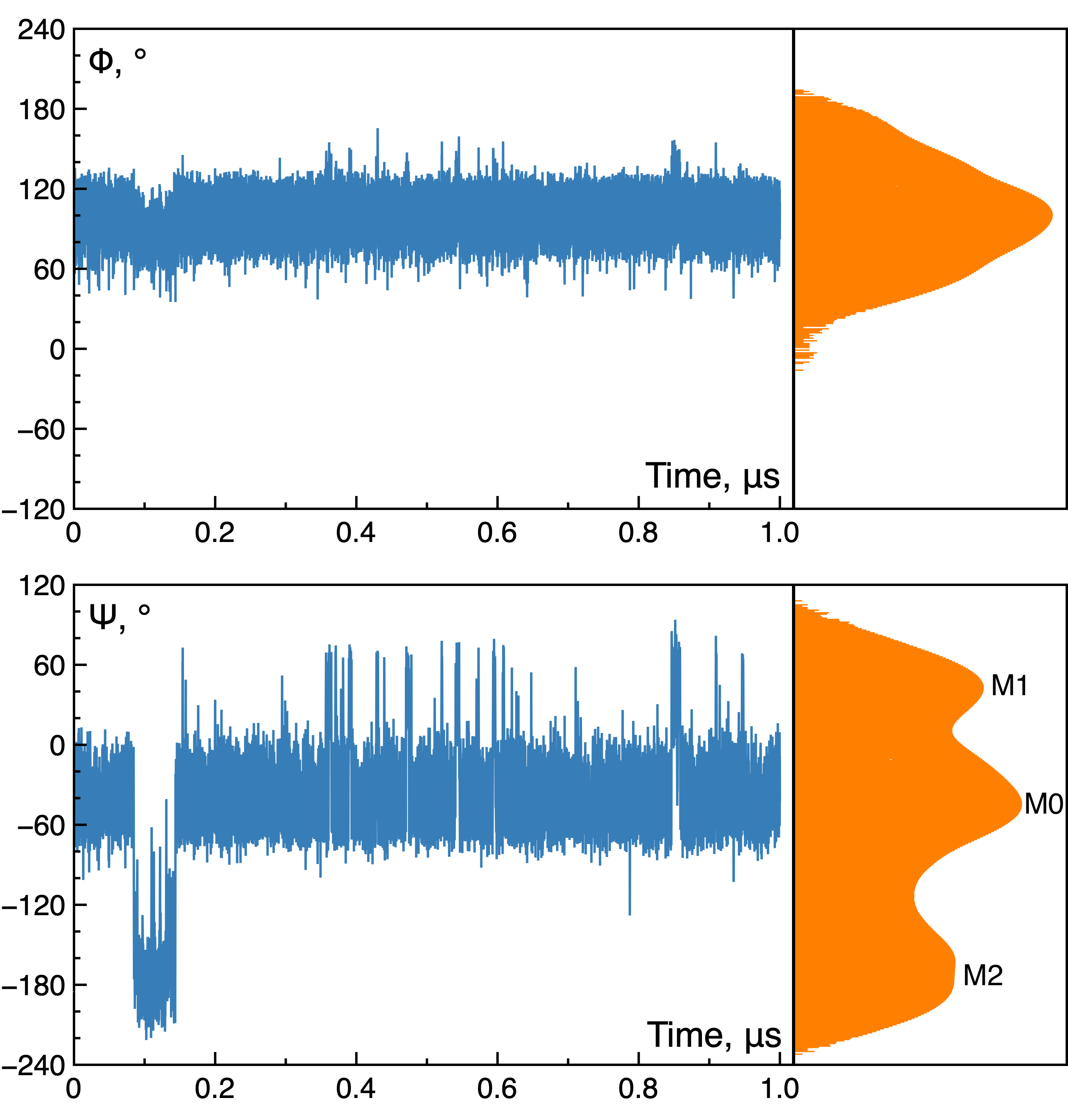}
\caption{The time evolution of the two dihedral angles, $\Phi$ and $\Psi$, describing the conformation of the glycosidic linkage in a sucrose molecule for one arbitrary molecule (for illustration) in the 50\% solution simulated with the OPLS-AA/1.14*CM1A-LBCC force field (left), and the distribution of these dihedral angles averaged over all sucrose molecules in the solution (right). The three modes in the $\Psi$ distribution correspond to the three frequently adopted conformations of the glycosidic linkage, M0, M1, and M2.}
\label{pic:phipsi}
\end{figure}

The left part of Fig.~\ref{pic:phipsi} shows the time evolution of the dihedral angles $\Phi$ and $\Psi$ for an arbitrary sucrose molecule in the 50\% solution using the OPLS-AA/1.14*CM1A-LBCC force field. The right part illustrates the distributions of both dihedral angles, averaged over all sucrose molecules in the system. The distribution of $\Phi$ is unimodal, whereas $\Psi$ exhibits three distinct modes, corresponding to the three most frequently adopted glycosidic linkage conformations: M0, M1, and M2. These conformations are discussed in the sections to follow.

\subsection{NMR J-coupling constants}
\begin{table}[h]
\centering
\resizebox{0.35\textwidth}{!}{%
\begin{tabular}{c|c|c|c|c}
\multirow{2}{*}{Atom Pair} & \multicolumn{3}{c|}{$J_{\mathrm{sim}}$, Hz} & \multirow{2}{*}{$J^{\mathrm{avg}}_{\mathrm{exp}}$, Hz} \\ \cline{2-4}
                           & FF 1          & FF 2         & FF 3         &                                                        \\ \hline
H1$_\mathrm{g}$ – H2$_\mathrm{g}$                  & 2.8           & 3.0          & 3.2          & 3.9                                                    \\
H2$_\mathrm{g}$ – H3$_\mathrm{g}$                  & 8.7           & 6.3          & 8.9          & 10.0                                                   \\
H3$_\mathrm{g}$ – H4$_\mathrm{g}$                  & 8.4           & 6.3          & 8.9          & 9.3                                                    \\
H4$_\mathrm{g}$ – H5$_\mathrm{g}$                  & 8.6           & 6.3          & 8.9          & 9.9                                                    \\
C3$_\mathrm{g}$ – H1$_\mathrm{g}$                  & 7.5           & 5.5          & 7.6          & 6.15                                                   \\
C4$_\mathrm{g}$ – H2$_\mathrm{g}$                  & 1.2           & 3.2          & 1.2          & 0.84                                                   \\
C5$_\mathrm{g}$ – H1$_\mathrm{g}$                  & 7.1           & 5.6          & 7.3          & 6.63                                                   \\ \hline
H3$_\mathrm{f}$ – H4$_\mathrm{f}$                  & 6.7           & 6.4          & 6.1          & 8.74                                                   \\
H4$_\mathrm{f}$ – H5$_\mathrm{f}$                  & 6.9           & 7.9          & 7.5          & 8.4                                                    \\
H5$_\mathrm{f}$ – H6R$_\mathrm{f}$                 & 3.5           & 1.6          & 2.6          & 3.7                                                    \\
H5$_\mathrm{f}$ – H6S$_\mathrm{f}$                 & 6.3           & 4.7          & 6.7          & 6.4                                                    \\
C3$_\mathrm{f}$ – C6$_\mathrm{f}$                  & 2.2           & 2.6          & 2.4          & 2.5                                                    \\
C1$_\mathrm{f}$ – H3$_\mathrm{f}$                  & 2.0           & 1.6          & 1.6          & 1.7                                                    \\
C6$_\mathrm{f}$ – H4$_\mathrm{f}$                  & 5.8           & 5.1          & 5.0          & 5.0                                                    \\ \hline
C2$_\mathrm{f}$ – C2$_\mathrm{g}$                  & 2.5           & 2.2          & 2.0          & 2.0                                                    \\
C2$_\mathrm{f}$ – H1$_\mathrm{g}$                  & 4.5           & 5.2          & 5.4          & 3.9                                                   
\end{tabular}
}
\caption{
{\color{black}
Calculated vicinal coupling constants ($J_\mathrm{sim}$) in sucrose molecule obtained using three force fields: GLYCAM06 (FF 1), OPLS-AA/1.14CM1A (FF 2), and OPLS-AA/1.14CM1A-LBCC (FF 3), along with the corresponding averaged experimental values ($J^{\mathrm{avg}}_{\mathrm{exp}}$). Atom pairs are divided into three groups: the first corresponds to the glucopyranose ring, the second to the fructofuranose ring, and the last group includes two atom pairs corresponding to dihedral angles through the glycosidic bond.
}
}\label{table:Jcoupling}
\end{table}

{\color{black}
To compare our simulations with the experimental dynamics of sucrose molecules in water, we calculated 16 vicinal J-coupling constants. The corresponding Karplus-like equations and experimental data used in this study are available in Table S1 of the supplementary material.

Table~\ref{table:Jcoupling} summarizes the J-coupling constants obtained from simulations using three force fields --- GLYCAM06 (FF 1), OPLS-AA/1.14CM1A (FF 2), and OPLS-AA/1.14CM1A-LBCC (FF 3) --- alongside reference values from NMR experiments. The data are divided into three groups. 

The first group includes seven atom pairs from the glucopyranose ring. The mean absolute errors (MAEs) for this group are 0.97~Hz (GLYCAM06), 2.18~Hz (OPLS-AA/1.14*CM1A), and 0.79~Hz (OPLS-AA/1.14*CM1A-LBCC). The poor agreement of the OPLS-AA/1.14*CM1A force field with experimental data is attributed to incorrect ring conformations, as discussed in Section~\ref{puckering}. In contrast, OPLS-AA/1.14*CM1A-LBCC shows better agreement than GLYCAM06, likely due to a lower population of the $^1$C$_4$ conformer of glucopyranose.

The second group consists of seven atom pairs from the fructofuranose ring. The corresponding MAEs are 0.77~Hz (GLYCAM06), 0.98~Hz (OPLS-AA/1.14CM1A), and~0.74 Hz (OPLS-AA/1.14*CM1A-LBCC). Again, OPLS-AA/1.14*CM1A-LBCC yields the best agreement with experimental data, while OPLS-AA/1.14*CM1A -- the worst. 

The final group contains two atom pairs corresponding to dihedral angles through the glycosidic bond. MAEs across all atom pairs are 0.82~Hz (GLYCAM06), 1.47~Hz (OPLS-AA/1.14*CM1A-LBCC), and 0.76~Hz (OPLS-AA/1.14*CM1A-LBCC). These results prove that the OPLS-AA/1.14*CM1A-LBCC force field provides the most accurate overall description of sucrose molecular dynamics in water. Detailed differences in the conformational preferences of sucrose in water for these force fields are discussed in the following sections.
}

\subsection{Ramachandran plots}

The Ramachandran plots were obtained for each system considered in this work. Figure S1 of the supplementary material presents these plots for simulations performed using the OPLS‑AA/1.14*CM1A‑LBCC force field. No significant differences were observed for sucrose solutions with mass fractions of 20\%, 30\%, and 50\%, each plot exhibiting three local minima corresponding to the sucrose conformers M0, M1, and M2. The dihedral angles $\Phi$ and $\Psi$ for each conformer at all concentrations are provided in Table S2 of the supplementary material. We verified that the free energy differences between the conformer pairs reached a stable plateau over the simulation time, confirming convergence (see Fig.~S2 of the supplementary material).

The lack of differences in the Ramachandran plots across the solutions indicates that the conformational preferences of sucrose in aqueous solution are unaffected by concentration. The positions of the local minima coincided within a margin of $3 \degree$, as shown in Table S2 of the supplementary material, in agreement with previous studies~\cite{xia2011sucrose1, xia2011sucrose2}.

The free energy surface remained nearly constant within a radius of $6 \degree$ around the point corresponding to conformer M2, as shown in the Ramachandran plots~(Fig.~S1 of the supplementary material) and the distribution of the $\Psi$ dihedral angle (Fig.~\ref{pic:phipsi}, bottom right) around $-178 \degree$. This observation suggests that the conformations of the sucrose molecule around the M2 conformer are equally probable and energetically equivalent. 

\begin{table*}
\centering
\begin{tabular}{|c|c|c|c|c|c|}
\hline
Force field               &                & M0       & M1      & M2       & M* \\ \hline
\multirow{2}{*}{OPLS-AA/1.14*CM1A-LBCC}     & $\Phi$, $\Psi$ & 102, -44 & 104, 44 & 86, -177 &    \\ \cline{2-6} 
                          & $F$, kcal/mol  & 0        & 1.85    & 3.04     &    \\ \hline
\multirow{2}{*}{GLYCAM06} & $\Phi$, $\Psi$ & 106, -58 & 88, 55  & 78, -168 &  72, -71 \\ \cline{2-6} 
                          & $F$, kcal/mol  & 0        & 1.87    & 1.49     & 0.29 \\ \hline
\end{tabular}%
\caption{Values of the dihedral angles $\Phi$ and $\Psi$ and the free energy
$F$ for the sucrose conformers identified in 50\% sucrose aqueous solutions for the OPLS-AA/1.14*CM1A-LBCC and GLYCAM06 force fields. The free energy is given relative to the global minimum, which corresponds to conformation M0 for both force fields.}
\label{table:newconf}
\end{table*}

In an earlier study, Immel et al.~\cite{immel1995molecular} examined sucrose conformation in vacuum using molecular mechanics with the PIMM88 force field. Two of the conformers identified in their study are located near the M0 and M2 conformers reported in this work, while the third one is shifted by $20 \degree$ from M1.

Xia and Case~\cite{xia2011sucrose1, xia2011sucrose2} performed extensive research on the conformation of the sucrose molecule in an aqueous solution using the GLYCAM06 force field and several water models. Four local minima in the Ramachandran plot were identified in their work. Three of them correspond to the M0, M1, and M2. One more sucrose conformer, labeled M* here, was found by Xia and Case, which can be related to the difference in force fields. They stated that this minimum exhibited the poorest agreement with experimental residual dipolar coupling and J-coupling data~\cite{xia2011sucrose1, xia2011sucrose2}. Our results show that OPLS‑AA/1.14*CM1A‑LBCC avoids this minimum. 

\begin{figure}
\centering
\includegraphics[width=1\linewidth]{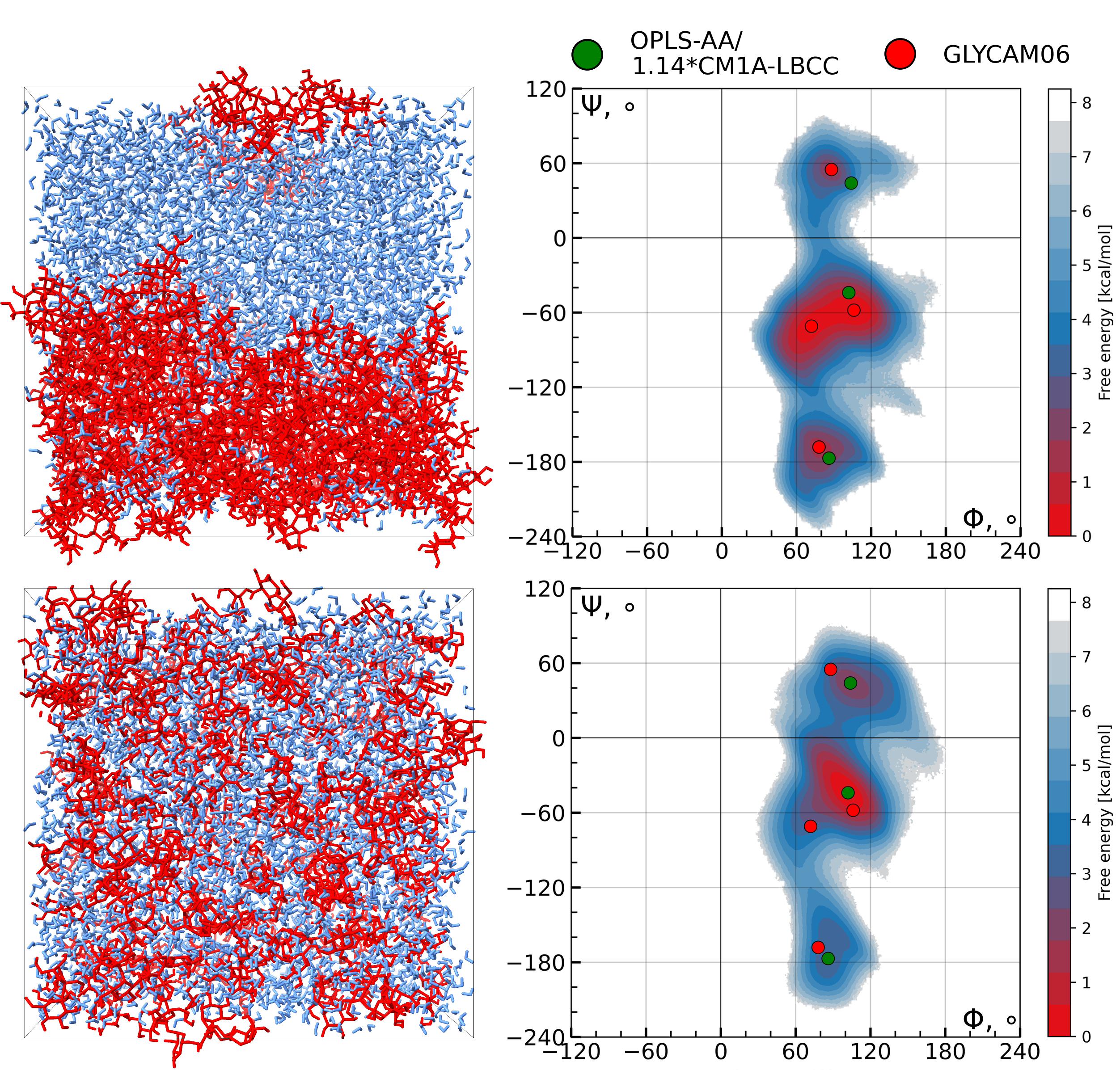}
\caption{Simulation box snapshots (left) and Ramachandran plots (right) for the aqueous solution simulations performed with the GLYCAM06 (top) and OPLS-AA/1.14*CM1A-LBCC (bottom) force fields. The sucrose and water molecules are represented in simulation boxes with red and blue colors, respectively. Local minima identified in this work are indicated by red dots for the GLYCAM06 and green dots for the OPLS‑AA/1.14*CM1A‑LBCC force field.}
\label{pic:newbig}
\end{figure}

For comparison, we also performed similar calculations using GLYCAM06 for a 50\% sucrose solution and obtained results matching to those reported by Xia and Case~\cite{xia2011sucrose1}. We likewise observed four conformers, including M*. The obtained Ramachandran plot is shown in right part of Fig.~\ref{pic:newbig}.
It is noteworthy that Xia and Case identified the conformers in a dilute solution, while our study considered the solution with the 50\% sucrose mass fraction. The similarity of the conformers identified in both cases is a further proof that sucrose’s conformational preferences are unaffected by concentration.

{The main problem of the GLYCAM06 force field is the overaggregation of sucrose molecules, which is not observed in the OPLS‑AA/1.14*CM1A‑LBCC force field, used in this work. This difference is illustrated in left part of Fig.~\ref{pic:newbig}, which shows snapshots of the 50\% sucrose solution simulation boxes for both force fields. For the GLYCAM06 force fields~(upper box) sucrose molecules formed a single large cluster, leaving most water molecules outside.}

In their first work~\cite{xia2011sucrose1}, Xia and Case considered only a single sucrose molecule in solution and still observed the M* minimum. This observation implies that M* is intrinsic to the GLYCAM06 force field rather than arising from overaggregation. We also performed two restrained simulations (with added torsion forces restraining the glycosidic linkage) of the 50\% sucrose solution to examine the opposite hypothesis that the M* conformer leads to clustering. However, the results did not support this hypothesis: in the GLYCAM06 simulation with the M0 conformer restrained, overaggregation still occurred, while in the OPLS‑AA/1.14*CM1A‑LBCC simulation with the M* conformer restrained, nucleation was not observed. Thus, the M* appearance is not connected with the overaggregation in the GLYCAM06 force field.

\subsection{Sucrose conformers}

\begin{figure*}%
\centering
\includegraphics[width=.9\linewidth]{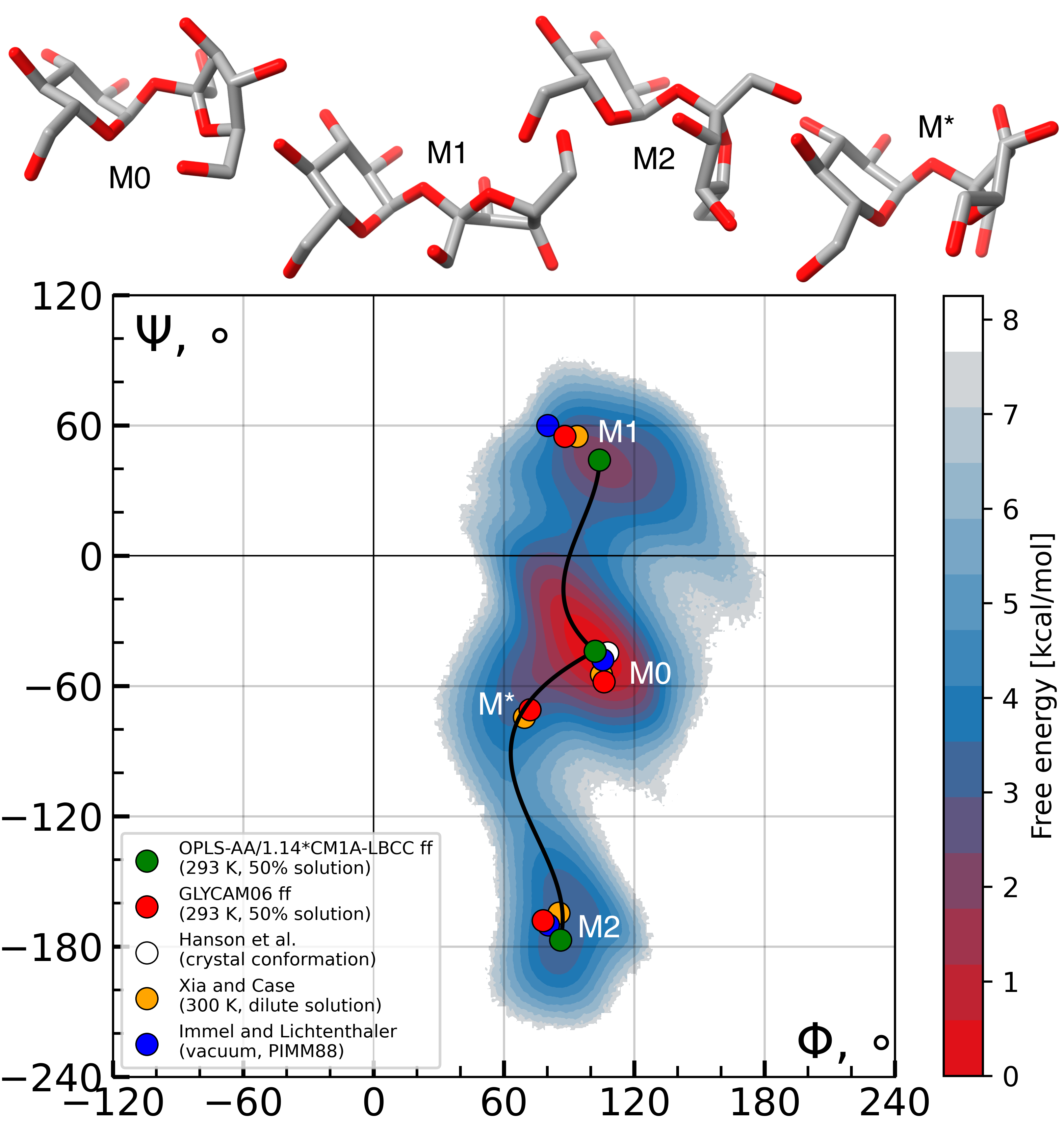}
\caption{Free energy surface of sucrose conformation as a function of the dihedral angles $\Phi$ and $\Psi$ in a solution with 50\% mass fraction of sucrose at 293~K obtained using OPLS-AA/1.14*CM1A-LBCC force field. The free energy surface is shifted to have a zero global minimum corresponding to conformer M0. The local minima, indicated by green dots, correspond to the sucrose conformers M0, M1, and M2. The visualizations of these conformers --- as well as of M*, which was identified only with GLYCAM06 force field -- are shown above the plot. The red dots denote the positions of conformers M0, M1, M2, and M* identified in this work using the GLYCAM06 force field.
The white dot marks the crystalline conformation of sucrose determined by X-ray crystallography by Hanson and colleagues~\cite{hanson1973sucrose}.
The orange dots denote stable conformations obtained using molecular dynamics with the GLYCAM06 force field for dilute sucrose solution at 300~K by Xia and Case~\cite{xia2011sucrose1}.
The blue dots indicate the minima obtained using the PIMM88 molecular mechanics method for sucrose in a vacuum by Immel and Lichtenthaler~\cite{immel1995molecular}.
The black curves represent transition pathways between the conformation pairs M2$\to$M0 and M0$\to$M1.}
\label{pic:9_ramabig}
\end{figure*}

Table~\ref{table:newconf} lists the positions of local minima identified in this work for both the GLYCAM06 and OPLS-AA/1.14*CM1A-LBCC force fields. Figure~\ref{pic:9_ramabig} shows the Ramachandran plot for the 50\% solution obtained using OPLS-AA/1.14*CM1A-LBCC force field. The locations of the local minima~(Table~\ref{table:newconf}) for OPLS-AA/1.14*CM1A-LBCC and GLYCAM06 force fields are marked with green and red dots, respectively. Sucrose conformers identified in other studies~\cite{hanson1973sucrose, immel1995molecular, xia2011sucrose1} are also shown with white, blue and orange dots.

Previous studies, along with the present one, identified multiple sucrose conformers in aqueous solution, specifically near conformers M0, M1, M2, and M*. The M0 conformer strongly resembles the crystal conformation revealed with neutron diffraction~\cite{brown1963sucrose} and X-ray crystallography~\cite{hanson1973sucrose}. In aqueous solution, however, the sucrose molecule has the additional conformers (M1, M2, and M*). This increased conformational variety is attributed to the interactions between sucrose and water, including the formation of hydrogen bonds between sucrose and water molecules and the disruption of intramolecular hydrogen bonds. 

{\color{black}
Several studies have previously examined the hydrogen bond formation between sucrose and water~\cite{Imberti2019, Olsson2020, Silva2022}. We also investigate the distribution of hydrogen bonds for the obtained conformers via the technique described in Section~\ref{h_bonds}. 

First of all, we have analyzed the intramolecular sucrose hydrogen bonds. For the M0 conformer, both crystal hydrogen bonds ($\mathrm{O1}_\mathrm{f}$–$\mathrm{HO1}_\mathrm{f} \cdots \mathrm{O2}_\mathrm{g}$ and $\mathrm{O6}_\mathrm{f}$–$\mathrm{HO6}_\mathrm{f} \cdots \mathrm{O5}_\mathrm{g}$) are formed during the simulation. For the M1 conformer, only the first one ($\mathrm{O1}_\mathrm{f}$–$\mathrm{HO1}_\mathrm{f} \cdots \mathrm{O2}_\mathrm{g}$) is formed, while in M2 none of them have been observed. It additionally indicates the difference between crystallic and M1, M2 conformers, observed in aqueous solution.

For the sucrose--water interactions, we have estimated the averaged number of hydrogen-bonded molecules for each conformer. The M0 conformer, which is closest one to the crystal conformation, has the lowest number -- 8.07, while other two conformers M1 and M2 have higher values -- 8.53 and 8.39, correspondingly. These values have a reasonable agreement with the results of study by Lerbret et al.~\cite{lerbret2005homogeneous}.
}

\begin{figure}%
\centering
\includegraphics[width=.95\linewidth]{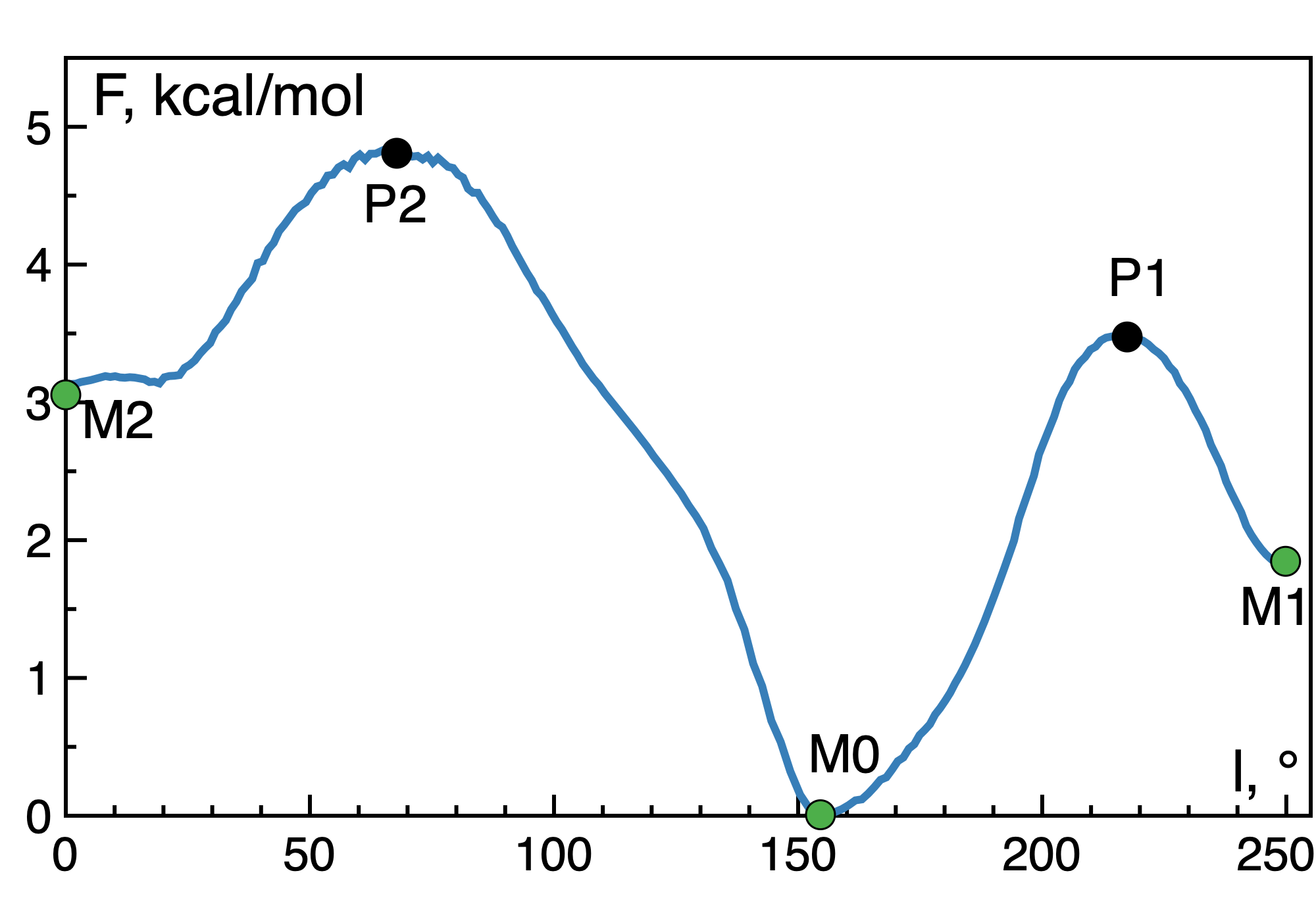}
\caption{Free energy profile $F$ along the transition pathways between conformers M2$\to$M0$\to$M1, shown in Fig.~\ref{pic:9_ramabig}. The x-axis represents the path length $l$ in degrees. The green dots indicate the sucrose conformers identified in this work, while the black dots, P1 (3.47~kcal/mol) and P2 (4.81~kcal/mol), mark the local maxima of the energy profile.}
\label{pic:10_energypaths}
\end{figure}

We investigated transitions between sucrose conformers, identified using the OPLS-AA/1.14*CM1A-LBCC force field. Transition pathways were located by approximating zero-gradient points between conformer pairs with third-order curves for the 50\% solution. The transition pathways are shown as black curves in Fig.~\ref{pic:9_ramabig}. Notably, this curve partially lies in the M* conformer’s zone. The free energy profiles along these pathways were obtained and shown in Fig.~\ref{pic:10_energypaths}. The M0, M1, and M2 minima, as well as the P1 and P2 maxima, are highlighted on Fig.~\ref{pic:10_energypaths}. These maxima were obtained by fitting the regions near the peaks with second-order curves.

These pathways are shown as black curves in Fig.~\ref{pic:9_ramabig}. Notably, part of one pathway overlaps the M region. The free energy profiles along these pathways appear in Fig.~\ref{pic:10_energypaths}, where the M0, M1, and M2 minima and the P1 and P2 maxima are labeled. These maxima were, again, determined by fitting the regions near the peaks with second-order curves.

\subsection{Conformers' lifetimes}
Using the all-atom trajectories from simulations performed with the OPLS‑AA/1.14*CM1A‑LBCC force field for solutions at three different sucrose concentrations, we computed the lifetimes of the sucrose conformers M0, M1, and M2. Conformers were identified using a geometric criterion based on the dihedral angles $\Phi$ and $\Psi$.

\begin{figure*}[t]
\centering
\includegraphics[width=.88\linewidth]{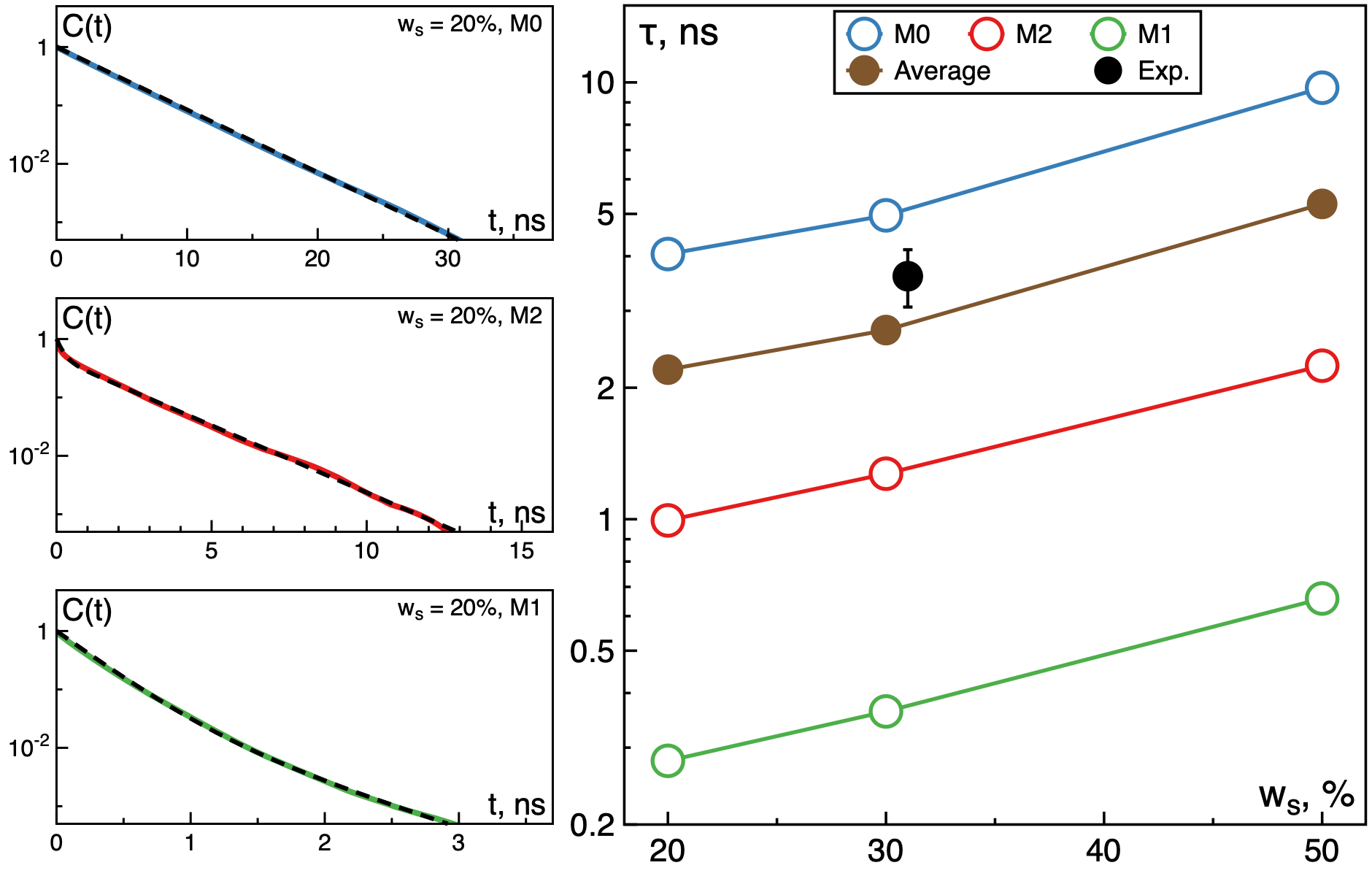}
\caption{(Left) Time autocorrelation functions $C(t)$ (Eq.~\eqref{eq:C}) for conformer M0 (top, blue color), M1 (bottom, green color), and M2 (center, red color) in a 50\% sucrose solution at 293~K, obtained using the OPLS-AA/1.14*CM1A-LBCC force field. The black curves represent the fits of the autocorrelation functions using Eq.~\eqref{eq:Cfit}. (Right) Concentration dependence of the characteristic lifetimes for M0 (blue circles), M1 (green circles), and M2 (red circles). Brown dots indicate the weighted average lifetime at each concentration, while the black dot represents the measured relaxation time associated with glycosidic linkage conformation changes reported by Behrends and Kaatze~\cite{kaatze2} for the 1~mol/L sucrose solution at 298~K.}
\label{pic:lifetimes}
\end{figure*}

{\color{black}
For each conformer (M0, M1, and M2), we calculated the time autocorrelation function $C(t)$, as defined in Eq.\eqref{eq:C}. The resulting autocorrelation functions for the 50\% sucrose solution are shown in the left panel of Fig.\ref{pic:lifetimes}, with blue, green, and red curves corresponding to M0, M1, and M2, respectively. Black dashed lines represent the double-exponential fits according to Eq.~\eqref{eq:Cfit}. Additional data for other concentrations are provided in Fig.~S3 of the supplementary material.

The autocorrelation function $C(t)$ for the M0 conformer is well described by a single exponential decay; accordingly, we performed the fit using Eq.~\eqref{eq:Cfit} with $A = 1$. In contrast, the M2 conformer displays a pronounced initial decay at short times ($t \sim 0.1$~ns) that deviates from a single-exponential behavior, while the M1 conformer also exhibits a clear biexponential decay in its autocorrelation function, indicating the presence of two distinct relaxation timescales.

Following the fitting procedure, we determined the characteristic lifetimes of the conformers, which are listed in Table~\ref{table:lifetimes} and shown in the right panel of Fig.~\ref{pic:lifetimes}. The conformer corresponding to the crystal structure (M0) consistently exhibited the longest characteristic lifetime at all concentrations, reflecting its superior stability relative to M1 and M2. The longer characteristic lifetime of M2 relative to M1 is explained by a higher potential energy barrier between M2 and M0, which reduces the frequency of transitions between them.}

\begin{table}
\centering
\begin{tabular}{|c|ccc|}
\hline
& $\tau$(M0), ns  & $\tau$(M1), ns & $\tau$(M2), ns \\ \hline
$w_s = 20\%$ & $4.05$ & $0.28$ & $0.99$ \\
$w_s = 30\%$ & $4.96$ & $0.36$ & $1.27$ \\
$w_s = 50\%$ & $9.72$ & $0.66$ & $2.25$ \\ \hline
\end{tabular}%
\caption{Characteristic lifetimes calculated using double exponential Eq.~\ref{eq:Cfit} for all conformers at different sucrose mass fractions in solution $w_s$: 20\%, 30\% and 50\%.}
\label{table:lifetimes}
\end{table}

The concentration dependence of the characteristic lifetimes is shown in the right part of Fig.~\ref{pic:lifetimes} for each conformer: M0~(blue symbols), M1~(green symbols), and M2~(red symbols). In addition, the brown symbols indicate the weighted average lifetime, calculated based on conformers' frequencies (i.e., the number of occurrences). A decrease in lifetime with rising sucrose concentration was observed for all conformers. 

This finding is in line with the works of Udo Kaatze and colleagues~\cite{kaatze1, kaatze2}, who experimentally investigated ultrasonic attenuation spectra for aqueous solutions of mono- and disaccharides and identified a disaccharide-specific relaxation process related to glycosidic linkage conformation changes. {\color{black}They assumed that these changes slow down with increasing viscosity and consequently with increasing sugar concentration in the aqueous solution.} For 1~mol/L sucrose solutions (sucrose mass fraction around $w_s = 31.3\%$) at 298~K, disaccharide-specific relaxation process was characterized by a relaxation time of $3.6 \pm 0.5$ ns, which is shown in Fig.~\ref{pic:lifetimes} by the black dot. 

{\color{black}\subsection{Fructofuranose ring puckering}\label{puckering1}
Other important degrees of conformational freedom in sucrose are the ring conformation changes. Accurately capturing the ring puckering distribution is critical for predicting the macroscopic hydrodynamic properties of carbohydrates~\cite{Plazinski2015, Sattellemacro2013}.

The fructofuranose ring conformation is characterized by the Altona--Sundaralingam parameters~\cite{Altona, Taha2013}, primarily the puckering phase $P$, which defines the spatial arrangement of atoms within the ring. We computed the free energy landscape of the ring as a function of phase across three force fields under investigation, with results presented in Fig.~\ref{pic:fes_fr}.

\begin{figure}[h]
\centering
\includegraphics[width=.9 \linewidth]{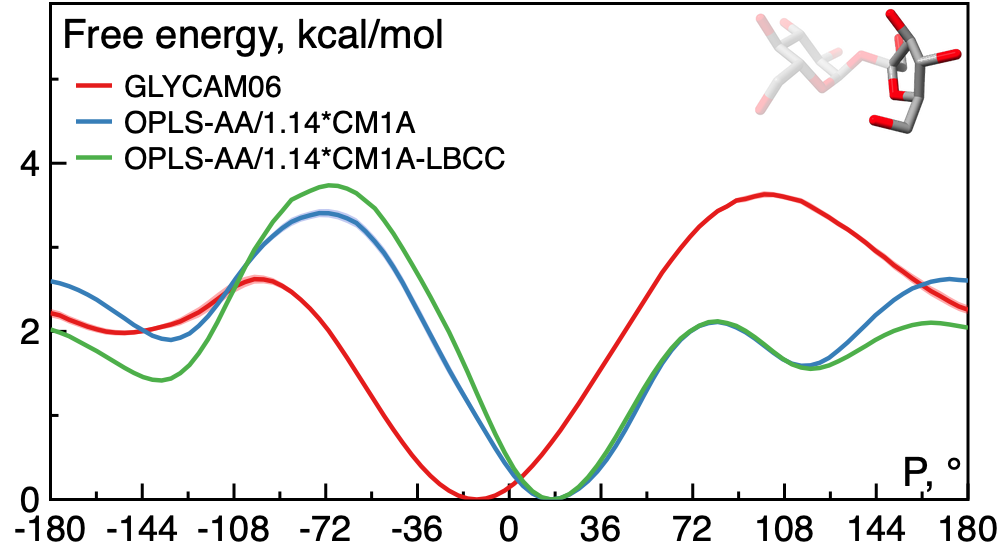}
\caption{{\color{black}
Free energy landscape of fructofuranose ring (highlighted in the insertion) as a function of Altona--Sundaralingam puckering phase $P$ for GLYCAM06 (red), OPLS-AA/1.14*CM1A (blue), and OPLS-AA/1.14*CM1A-LBCC (green) force fields in 50\% solutions at 293K. 
}}
\label{pic:fes_fr}
\end{figure}

For the GLYCAM06 force field, only two minima near $-18\degree$ (global minima) and $-144\degree$ were identified. These positions agree with the work of Xia and Case~\cite{xia2011sucrose1}, which reported probability peaks at $-5\degree$ and $-150\degree$. For the OPLS-AA/1.14*CM1A force fields, three local minima were observed near $18\degree$ (global minima), $126\degree$, and $-126\degree$.

The peaks near $0\degree$ represent the global minima for all three force fields and correspond to the crystal conformation of the fructofuranose ring~\cite{Taha2013}. However, solution-state NMR data and computational studies consistently demonstrate greater ring flexibility in aqueous environments, predominantly within the northern hemisphere of the pseudorotation wheel~\cite{duker1993,freedberg2002,xia2011sucrose1}.
}

\subsection{Glucopyranose ring puckering}\label{puckering}

We investigated glucopyranose ring puckering in the 50\% sucrose solution using the Cremer--Pople analysis~\cite{cremer1975, ring2021} for three force fields: GLYCAM06, OPLS-AA/1.14*CM1A-LBCC, and OPLS-AA/1.14*CM1A. The $^4\mathrm{C}_1$ ($^1\mathrm{C}_4$) conformation was defined as instances when the Cremer--Pople $\Theta$ angle was less than 30\degree (more than 150\degree). In all cases, the starting conformation of sucrose was the crystal structure~\cite{hanson1973sucrose}, featuring the $^4\mathrm{C}_1$ chair --- the conventional glucopyranose ring conformation in the aqueous solution. However, during the simulations, other conformations, such as the $^1\mathrm{C}_4$ chair and non-chair ones, were also observed across all force fields.

\begin{figure}[h!]
\centering
\includegraphics[width=.9 \linewidth]{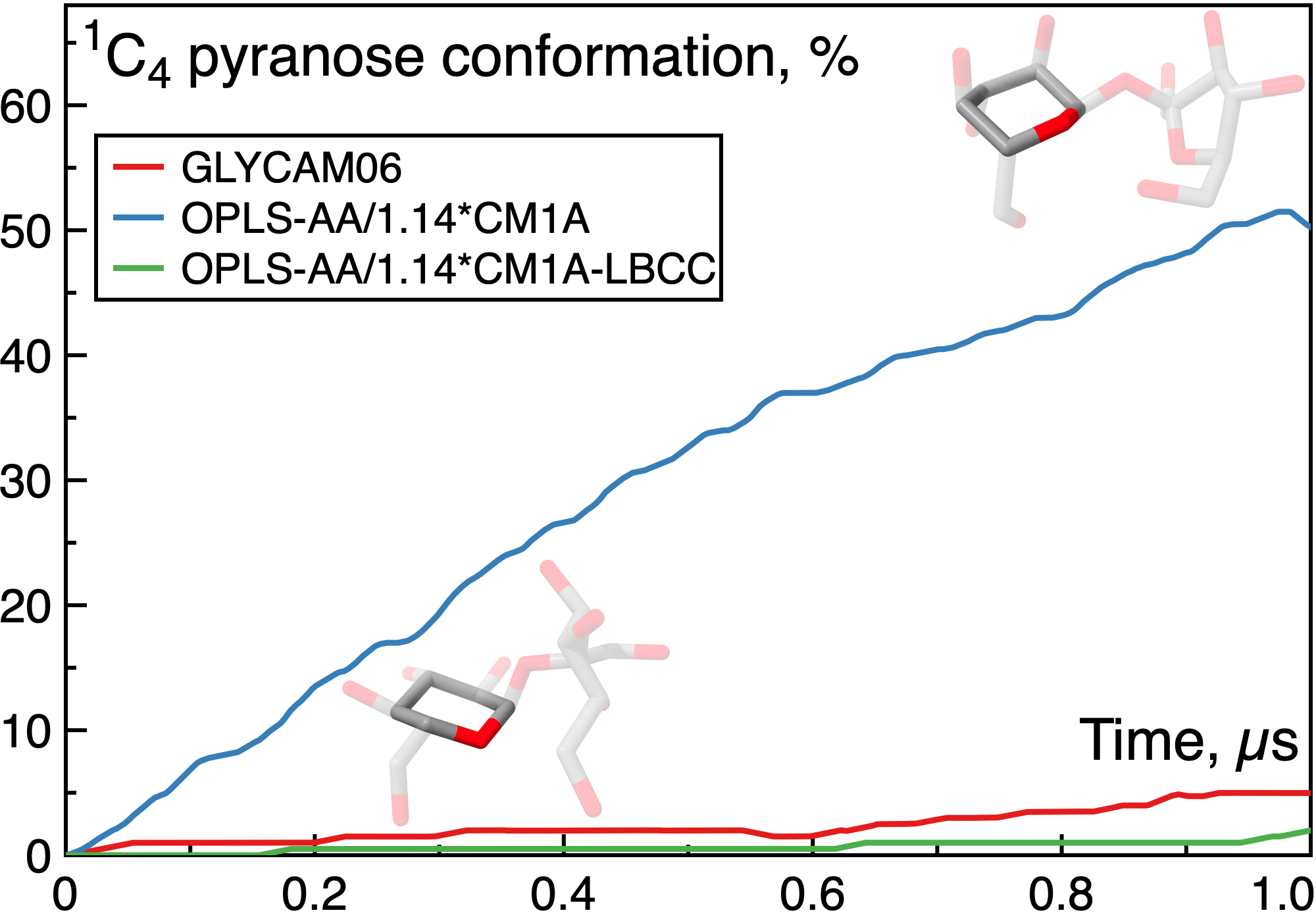}
\caption{Time evolution of the percentage of the unconventional $^1\mathrm{C}_4$ glucopyranose ring conformation for the GLYCAM06 (red), OPLS-AA/1.14*CM1A (blue), and OPLS-AA/1.14*CM1A-LBCC (green) force fields for 50\% solutions at 293~K. A moving average smoothing was applied to all curves. Two sucrose molecule visualizations are presented: the one with the $^4\mathrm{C}_1$ glucopyranose ring conformation appears on the left, and the one with the $^1\mathrm{C}_4$ on the right.}
\label{pic:pyrring}
\end{figure}

For the GLYCAM06 and OPLS-AA/1.14*CM1A-LBCC force fields, the $^4\mathrm{C}_1$ conformation predominated, representing approximately 94–96\% of the observed conformations. In contrast, for OPLS-AA/1.14*CM1A the population of the $^1\mathrm{C}_4$ conformation increased significantly, suggesting that $^1\mathrm{C}_4$ is more stable and energetically favored for this force field. {\color{black}This contradicts available experimental data~\cite{Deegbey, reeves1950shape, franks1989isomeric}}. Figure~\ref{pic:pyrring} shows the time evolution of the $^1\mathrm{C}_4$ conformation.

{\color{black}
Differences in the $^1\mathrm{C}_4$ conformer populations among the force fields result in variations in the J-coupling constants calculated for the glucopyranose ring (first group in Table~\ref{table:Jcoupling}). MAE between the experimental and simulated values decreases as the $^1\mathrm{C}_4$ population decreases across these force fields. This trend confirms that the preferred conformation of the glucopyranose ring in solution is $^4\mathrm{C}_1$.
}

Additionally, we computed Ramachandran plots using only sucrose molecules in the $^1\mathrm{C}_4$ glucopyranose conformation extracted from trajectories obtained with the OPLS‑AA/1.14*CM1A‑LBCC and GLYCAM06 force fields~(see Fig.~S4 of the supplementary material). For the OPLS‑AA/1.14*CM1A‑LBCC force field, we observed the M* minimum, which had not been observed in the previous analysis of the entire system~(Fig.~\ref{pic:9_ramabig}). For the GLYCAM06 force field, the M* conformer dominated, while the M0 minimum was absent. These findings indicate that the $^1\mathrm{C}_4$ glucopyranose ring conformation stabilizes the M* conformer. Because the OPLS‑AA/1.14*CM1A force field yields a higher population of the $^1\mathrm{C}_4$ glucopyranose ring conformation, the M* conformer is consequently observed, consistent with our previous findings~\cite{deshchenya2022molecular}.

It should be noted that convergence was not achieved within the simulation time for any of the force fields, likely because the conformational transitions occur on a microsecond timescale. Longer trajectories are necessary for a comprehensive analysis using conventional molecular dynamics.

{\color{black}
This problem is traditionally addressed for carbohydrates using metadynamics~\cite{Spiwok2010,Alibay2019,Liao2025}. However, computational studies specifically targeting glucopyranose ring in sucrose remain scarce. In this work, we employed well-tempered metadynamics via the OpenMM-PLUMED interface~\cite{Plumed2} for a dilute sucrose solution (1000 water molecules in the simulation cell).

The Cremer--Pople Cartesian coordinates served as collective variables for the bias potential. Gaussian hills were deposited every 1~ps, with widths of 0.05~\AA, 0.05~\AA, and 0.03~\AA\, along the three collective variable dimensions, an initial height of 0.3~kcal/mol, and a bias factor of 10.

We utilized a reweighting procedure to obtain free energy landscapes as a function of $\Theta$ and estimate the free energy difference between the $^4C_1$ and $^1C_4$ conformations for the glucopyranose ring in sucrose: 1.3~kcal/mol for GLYCAM06 and 0.8~kcal/mol for OPLS-AA/1.14*CM1A-LBCC. For comparison, Alibay and Bryce reported a $^4C_1 \rightarrow\, ^1C_4$ chair-inversion free energy value of 0.2 kcal/mol for $\alpha$-D-glucopyranose using the GLYCAM06 force field. This computational value is significantly lower than the experimental one of 4.2 kcal/mol determined by Angyal et al.~\cite{Angyal1968}.

Figure~\ref{pic:fes_gl} presents the resulting landscapes for the three force fields. For OPLS-AA/1.14*CM1A, the global minimum occurs at $\Theta \approx 180\degree$, indicating that the $^1C_4$ conformer is the most stable. In contrast, the other two force fields stabilize the $^4C_1$ conformer of the glucopyranose ring in sucrose.

\begin{figure}[h!]
\centering
\includegraphics[width=.9 \linewidth]{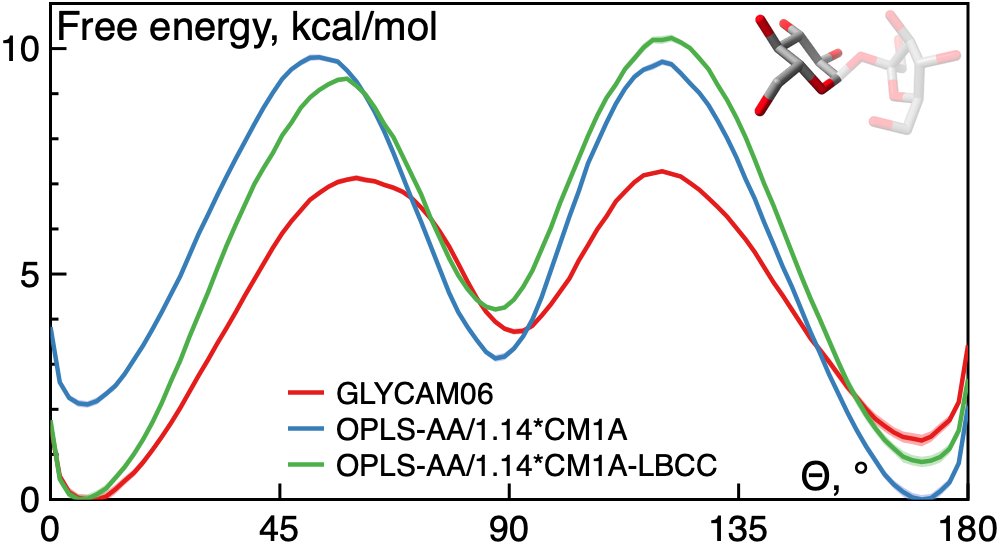}
\caption{{\color{black}
Free energy landscape for glucopyranose ring (highlighted in the insertion) as a function of Cremer--Pople puckering phase $\Theta$ for the GLYCAM06 (red), OPLS-AA/1.14*CM1A (blue), and OPLS-AA/1.14*CM1A-LBCC (green) force fields for 50\% solutions at 293K.
}}
\label{pic:fes_gl}
\end{figure}
}

\section{Conclusion}
In this work, we carried out microsecond molecular dynamics simulations to explore the conformational behavior of sucrose in aqueous solutions across a range of concentrations (20\%, 30\%, and 50\% by mass). We focused on glycosidic linkage conformers and their lifetimes, supported by comparisons with the J-coupling constant obtained by NMR, experimental ultrasonic spectra, and previous computational works. The comparison of two force fields, OPLS-AA/1.14*CM1A-LBCC and GLYCAM06, were conducted in terms of conformers and sucrose molecule dynamics averaged behavior. Furthermore, we analyzed the influence of glucopyranose ring puckering on the glycosidic linkage conformation for these two force fields, as well as for the OPLS-AA/1.14*CM1A force field.

Ramachandran plot analysis revealed three principal glycosidic linkage conformers (M0, M1, and M2) using OPLS-AA/1.14*CM1A-LBCC, whereas an additional conformer (M*) emerged under GLYCAM06. Xia and Case previously stated that this extra conformer M* has weakest agreement with NMR experimental data. Our J-coupling constant analysis further supported the accuracy of OPLS-AA/1.14*CM1A-LBCC, as it yielded lower MAE value relative to GLYCAM06 and OPLS-AA/1.14*CM1A.

We also quantified the conformers lifetimes for OPLS-AA/1.14*CM1A-LBCC force field at different sucrose concentrations. The M0 conformer --- closely matching the crystalline structure --- showed the longest characteristic lifetime (4.05--9.72~ns), while M2 (0.99--2.25~ns) exhibited longer lifetime than M1 (0.28--0.66~ns), consistent with a higher free energy barrier separating M2 from M0. The increase in characteristic lifetime with increasing concentration was observed for all conformers, which is consistent with the findings of previous experimental works. The concentration dependence of the weighted average lifetime was calculated and found to be in strong agreement with the experimental value (3.6~ns) obtained from ultrasonic spectra.

{\color{black}Analysis of fructofuranose ring puckering using the Altona--Sundaralingam parameters reveals a prevalent puckering phase near $P \approx 0\degree$ for all three force fields, corresponding to the crystal conformation. Additionally, the OPLS force fields exhibit two additional local minima at $P \approx -126\degree$ and $P \approx 126\degree$, while GLYCAM06 shows a one more local minimum at $P \approx -144\degree$.
}

Moreover, we investigated glucopyranose ring puckering through Cremer--Pople analysis, showing that all three force fields sample both $^4\mathrm{C}_1$ and $^1\mathrm{C}_4$ conformations differently. While $^4\mathrm{C}_1$ dominates in GLYCAM06 and OPLS‑AA/1.14*CM1A‑LBCC, the unconventional $^1\mathrm{C}_4$ conformation is more prevalent in OPLS‑AA/1.14*CM1A. {\color{black}This underscores a limitation of the OPLS-AA/1.14*CM1A force field, making it unsuitable for accurately modeling sucrose. 
Metadynamics simulations confirm these observations and provide $^4C_1 \rightarrow\, ^1C_4$ chair-inversion free energy value of 1.3~kcal/mol for GLYCAM06 and 0.8~kcal/mol for OPLS-AA/1.14*CM1A-LBCC.} We also observed that $^1\mathrm{C}_4$ can stabilize the M* glycosidic linkage conformer, highlighting the interplay between glucopyranose ring puckering and glycosidic linkage conformation.

Overall, these findings demonstrate that the OPLS-AA/1.14*CM1A-LBCC force field is a robust model for investigating sucrose aqueous solutions, accurately capturing both structural and dynamical properties~\cite{deshchenya2022molecular} while avoiding overaggregation.

\section*{Supplementary Material}
{\color{black}See \href{https://figshare.com/articles/figure/Supporting_Information/29422259?file=55709522}{supplementary material} for the additional information about free energy surfaces at various concentrations, time evolution of free energy differences between conformer pairs, all the time autocorrelation functions $C(t)$ at different concentrations, J-coupling constants and values of dihedral angles for each conformer.
}

\begin{acknowledgments}
V.I.D. is supported by the RSF project 20-71-10127 (optimization of computational performance for getting long MD trajectories). The work is supported by the Ministry of Science and Higher Education of the Russian Federation (Agreement No.  075-03-2025-662 17.01.2025) (force fields comparison for aqueous solutions, calculation methods). The calculations are performed on the \textit{Soft Cluster} of the Center for Computational Physics MIPT.
\end{acknowledgments}

\section*{Data Availability Statement}

The data that support the findings of this study are available from the corresponding author upon reasonable request.

\bibliography{main}

\end{document}